\documentclass[%
 reprint,
superscriptaddress,
nobibnotes,
 amsmath,amssymb,
 longbibliography,
 aps,
 prl,
]{revtex4-2}

\usepackage{graphicx}
\usepackage{dcolumn}
\usepackage{bm}
\usepackage{xcolor}
\usepackage{hyperref}
\usepackage{accents}

\usepackage{upgreek}

\begin{document}

\title{Precise and accurate nuclear electric quadrupole moment of $^{223}$Ra from molecular spectroscopy}

\author{C.~M.~Fajardo-Zambrano}
\thanks{These authors contributed equally to this work. Contact authors: carlos.fajardo@kuleuven.be, m.athkak@cern.ch}
\affiliation{KU Leuven, Instituut voor Kern- en Stralingsfysica, Leuven, Belgium}

\author{M.~Athanasakis-Kaklamanakis}
\thanks{These authors contributed equally to this work. Contact authors: carlos.fajardo@kuleuven.be, m.athkak@cern.ch}
\affiliation{KU Leuven, Instituut voor Kern- en Stralingsfysica, Leuven, Belgium}
\affiliation{Experimental Physics Department, CERN, Geneva, Switzerland}
\affiliation{JILA and University of Colorado, Boulder, CO 80309, USA}


\author{L.~Lalanne}
\affiliation{Experimental Physics Department, CERN, Geneva, Switzerland}
\affiliation{Université Paris-Saclay, CNRS/IN2P3, IJCLab, Orsay, 91405, France}

\author{J. R. Reilly}
\affiliation{Department of Physics and Astronomy, The University of Manchester, Manchester, UK}

\author{O.~Ahmad}
\affiliation{KU Leuven, Instituut voor Kern- en Stralingsfysica, Leuven, Belgium}

\author{M.~Au}
\affiliation{Systems Department, CERN, Geneva, Switzerland}

\author{S.~W.~Bai}
\affiliation{School of Physics and State Key Laboratory of Nuclear Physics and Technology, Peking University, Beijing, China}

\author{J.~Berbalk}
\affiliation{KU Leuven, Instituut voor Kern- en Stralingsfysica, Leuven, Belgium}
\affiliation{Systems Department, CERN, Geneva, Switzerland}

\author{C.~Bernerd}
\affiliation{Systems Department, CERN, Geneva, Switzerland}

\author{A.~A.~Breier}
\affiliation{Institut f\"ur Physik und Astronomie, Technische Universit\"at Berlin, Berlin, Germany}
\affiliation{Laboratory for Astrophysics, Institute of Physics, University of Kassel, Kassel, Germany}

\author{K.~Chrysalidis}
\affiliation{Systems Department, CERN, Geneva, Switzerland}

\author{T.~E.~Cocolios}
\affiliation{KU Leuven, Instituut voor Kern- en Stralingsfysica, Leuven, Belgium}

\author{R.~P.~de~Groote}
\affiliation{KU Leuven, Instituut voor Kern- en Stralingsfysica, Leuven, Belgium}

\author{K.~T.~Flanagan}
\affiliation{Department of Physics and Astronomy, The University of Manchester, Manchester, UK}
\affiliation{Photon Science Institute, The University of Manchester, Manchester, UK}

\author{S.~Franchoo}
\affiliation{Université Paris-Saclay, CNRS/IN2P3, IJCLab, Orsay, 91405, France}

\author{R.~F.~Garcia~Ruiz}
\affiliation{Department of Physics, Massachusetts Institute of Technology, Cambridge, MA, USA}
\affiliation{Laboratory for Nuclear Science, Massachusetts Institute of Technology and MIT-Harvard Center for Ultracold Atoms, Cambridge, MA, USA}

\author{R.~Heinke}
\affiliation{Systems Department, CERN, Geneva, Switzerland}

\author{P.~Imgram}
\affiliation{KU Leuven, Instituut voor Kern- en Stralingsfysica, Leuven, Belgium}

\author{A.~Koszor\'us}
\affiliation{KU Leuven, Instituut voor Kern- en Stralingsfysica, Leuven, Belgium}

\author{P.~Lass\`egues}
\affiliation{KU Leuven, Instituut voor Kern- en Stralingsfysica, Leuven, Belgium}

\author{J.~Lim}
\affiliation{Centre for Cold Matter, Blackett Laboratory, Imperial College London, London, UK}

\author{Y.~C.~Liu}
\affiliation{School of Physics and State Key Laboratory of Nuclear Physics and Technology, Peking University, Beijing, China}

\author{K.~M.~Lynch}
\affiliation{Department of Physics and Astronomy, The University of Manchester, Manchester, UK}

\author{A.~McGlone}
\affiliation{Department of Physics and Astronomy, The University of Manchester, Manchester, UK}

\author{W.~C.~Mei}
\affiliation{School of Physics and State Key Laboratory of Nuclear Physics and Technology, Peking University, Beijing, China}

\author{G.~Neyens}
 \email{gerda.neyens@kuleuven.be}
\affiliation{KU Leuven, Instituut voor Kern- en Stralingsfysica, Leuven, Belgium}

\author{L.~Nies}
\affiliation{Experimental Physics Department, CERN, Geneva, Switzerland}

\author{A. Raggio}
\affiliation{Department of Physics, University of Jyv\"askyl\"a, Jyväskylä, Finland}

\author{S. Rothe}
\affiliation{Systems Department, CERN, Geneva, Switzerland}

\author{B. van den Borne}
\affiliation{KU Leuven, Instituut voor Kern- en Stralingsfysica, Leuven, Belgium}

\author{J. Warbinek}
\affiliation{GSI Helmholtzzentrum f\"ur Schwerionenforschung GmbH, Darmstadt, Germany}
\affiliation{Department of Chemistry - TRIGA Site, Johannes Gutenberg-Universit\"at Mainz, Mainz, Germany}

\author{J. Wessolek}
\affiliation{Systems Department, CERN, Geneva, Switzerland}
\affiliation{Department of Physics and Astronomy, The University of Manchester, Manchester, UK}

\author{X. F. Yang}
\affiliation{School of Physics and State Key Laboratory of Nuclear Physics and Technology, Peking University, Beijing, China}

\date{\today}

\begin{abstract}
Nuclear electric quadrupole moments of unstable isotopes are commonly extracted from laser spectroscopy of the atomic hyperfine structure, where the accuracy is limited to the several-percent level by the available electric-field-gradient (EFG) calculations. In this study, we determine the spectroscopic quadrupole moment of $^{223}$Ra (half-life $t_{1/2}= 11.4$\,d) from hyperfine-resolved collinear resonance ionization spectroscopy of the radioactive molecule $^{223}$Ra$^{19}$F. The $X\,^2\Sigma^+$ electronic ground state of RaF provides a large axial EFG at the radium nucleus that can be calculated with relativistic coupled-cluster theory to sub-percent accuracy. Together with a three-fold improvement in spectral resolution over previous fast-beam measurements of radioactive molecules, this yields $Q_s = +1.211(7)_{\rm{stat}}(2)_{\rm{syst}}(5)_{\rm{theo}}$\,$e$b for $^{223}$Ra with a total uncertainty of 0.73\%, approximately three times better than atomic and ionic determinations. This result demonstrates that molecular spectroscopy can provide access to the electromagnetic structure of unstable nuclei with a precision competitive with, and in favorable cases better than, atomic spectroscopy, and may thus be envisioned for studies beyond the limitations of current radioactive-atom experiments.

\end{abstract}

\maketitle

\paragraph{\label{sec:introduction}Introduction---}The nuclear electric quadrupole moment quantifies the deviation of the nuclear charge distribution from spherical symmetry, thus being one of the most direct probes of nuclear deformation~\cite{neyens2003nuclear}. Together with nuclear spins, magnetic dipole moments, and changes in mean-squared charge radii, quadrupole moments have been measured extensively across the nuclear chart using laser spectroscopy, providing nuclear-model-independent access to the structure of nuclear ground and isomeric states far from stability~\cite{yang2023laser,nortershauser2022nuclear,koszorus2024laser}.

For the radium isotopes with mass numbers around $A\approx 220$\,--\,$229$, precise quadrupole moments are of particular interest. These nuclei lie in a region of stable octupole deformation, where the interplay of quadrupole and octupole collectivity~\cite{Robledo2013} gives rise to static nuclear shapes that break reflection symmetry~\cite{Butler1996}, as opposed to soft quadrupole and octupole shape vibrations~\cite{Butler2019Radon}, while also showing an inversion of the odd-even staggering in charge radii~\cite{Verstraelen2019,lynch2018laser}. In addition to its role in studies of the rich many-body physics manifesting in atomic nuclei~\cite{bohr1977nuclear,Butler1996,verney2025history}, deformation also enhances the symmetry-violating nuclear Schiff moment and, with it, the sensitivity of radium-containing atoms and molecules to potential physics beyond the Standard Model of particle physics~\cite{arrowsmith2023opportunities}.


Short-lived radioactive isotopes must be produced and studied \textit{in situ} at specialized radioactive ion beam (RIB) facilities, where nuclear reactions between energetic projectile beams and target nuclei create the nuclides of interest. In such facilities, the production of an intense and pure ion beam of a given isotope suitable for laser spectroscopy depends strongly on the given element's chemical behavior. In the isotope separation online method~\cite{lindroos2004review}, nuclides are produced within thick solid or liquid targets and must diffuse through the material and desorb from the surface before ionization~\cite{koster2002intense}. Isotopes of refractory elements are therefore generally inaccessible, as they require very high operating temperatures that jeopardize the integrity of the target material. Alternatively, nuclides produced in thin targets or by in-flight fragmentation must be stopped in a gas cell before being studied with laser spectroscopy~\cite{cheal2012collinear,blumenfeld2013facilities}. Within the gas cell, isotopes of reactive elements tend to form complex molecules with impurities in the gas, preventing them from emerging as atomic ion beams.

A promising approach to access the electromagnetic structure of refractory and reactive radioactive isotopes via laser spectroscopy is to study them as part of a molecule. The molecular hyperfine structure can be used to extract the nuclear spin and electromagnetic moments~\cite{BrownCarringtonBook}, as well as information on the nuclear charge radius~\cite{athanasakis2023king}. Rapid progress in the study of short-lived radioactive molecules has been seen in the last decade with the study of radium monofluoride $^{225,226}$Ra$^{19}$F~\cite{GarciaRuiz2020,udrescu2024precision,wilkins2025observation,athanasakis2025electron,athkak2024}, radium monohydroxide $^{226}$Ra$^{16}$O$^{1,2}$H~\cite{conn2025productionspectroscopycoldradioactive}, and actinium monofluoride $^{227}$Ac$^{19}$F~\cite{athkak2025AcF}. Recently, the study of $^{225}$Ra$^{19}$F allowed the extraction of the magnetic dipole moment and demonstrated that molecular spectroscopy can probe hyperfine anomalies and the radial distribution of nuclear magnetization~\cite{wilkins2025observation}. While molecular spectroscopy has long provided some of the most accurate quadrupole moments of stable nuclei~\cite{Pyykko2018}, the electric quadrupole hyperfine term in RaF has not yet been studied, as the $I=1/2$ spin of $^{225}$Ra does not permit the observation of an electric quadrupole moment. Thus, molecular measurements of electric quadrupole moments of short-lived isotopes have not been reported to date.


Here, we present hyperfine-resolved collinear resonance ionization spectroscopy of $^{223}$Ra$^{19}$F ($I=3/2$, $t_{1/2}=11.4$\,d) at CERN-ISOLDE, achieving a three-fold improvement in spectral resolution compared to previous fast-beam measurements of RaF \cite{udrescu2024precision}. By combining the measured quadrupole hyperfine constant of the $X\,^2\Sigma^+$ state with the calculated electric field gradient (EFG) at the radium nucleus, the value $Q_s(^{223}\rm{Ra}) = +1.211(7)_{\rm{stat}}(2)_{\rm{syst}}(5)_{\rm{theo}}$\,$e$b is determined. The total uncertainty is reduced by a factor of three relative to the previously evaluated value, $Q_s = 1.21(3)$\,$e$b \cite{pyykko2008year}. These results demonstrate that, when combined with advances in \textit{ab initio} quantum chemistry calculations, molecular spectroscopy can match, and in some cases exceed, the precision of atomic studies for short-lived isotopes. The prospects for using molecular spectroscopy to study the properties of radioactive isotopes that have so far eluded spectroscopic studies are also discussed.

\paragraph{Methods---}Hyperfine-resolved laser spectroscopy of $^{223}$Ra$^{19}$F was performed using the collinear resonance ionization spectroscopy (CRIS) experiment at the ISOLDE RIB facility at CERN. The production mechanism and spectroscopic apparatus have been described in detail in Refs.~\cite{udrescu2024precision,wilkins2025observation,athanasakis2025electron} and are briefly reviewed here.


$^{223}$Ra nuclides were produced by impinging pulses of 1.4-GeV protons onto a thick UC$_{x}$ target kept slightly above room temperature. Radionuclides were later released by heating the target container to $2000\,^\circ$C, and $^{223}$Ra$^{19}$F molecules were formed by injecting CF$_{4}$ into the container. The molecules were ionized by collisions with the walls of a tantalum capillary maintained at $2000\,^\circ$C, after which the molecular ions were accelerated to $30\,$keV and separated from contaminant species using two magnetic dipole separators in series. The mass-selected molecular ion beam was cooled to room temperature in a gas-filled radiofrequency quadrupolar cooler-buncher~\cite{Mane2009}, which released the continuous ion beam in the form of ion bunches with a temporal width of 5\,$\upmu$s once every 10\,ms to the CRIS experiment for study.

\begin{figure*}
\includegraphics[width=0.99\textwidth]{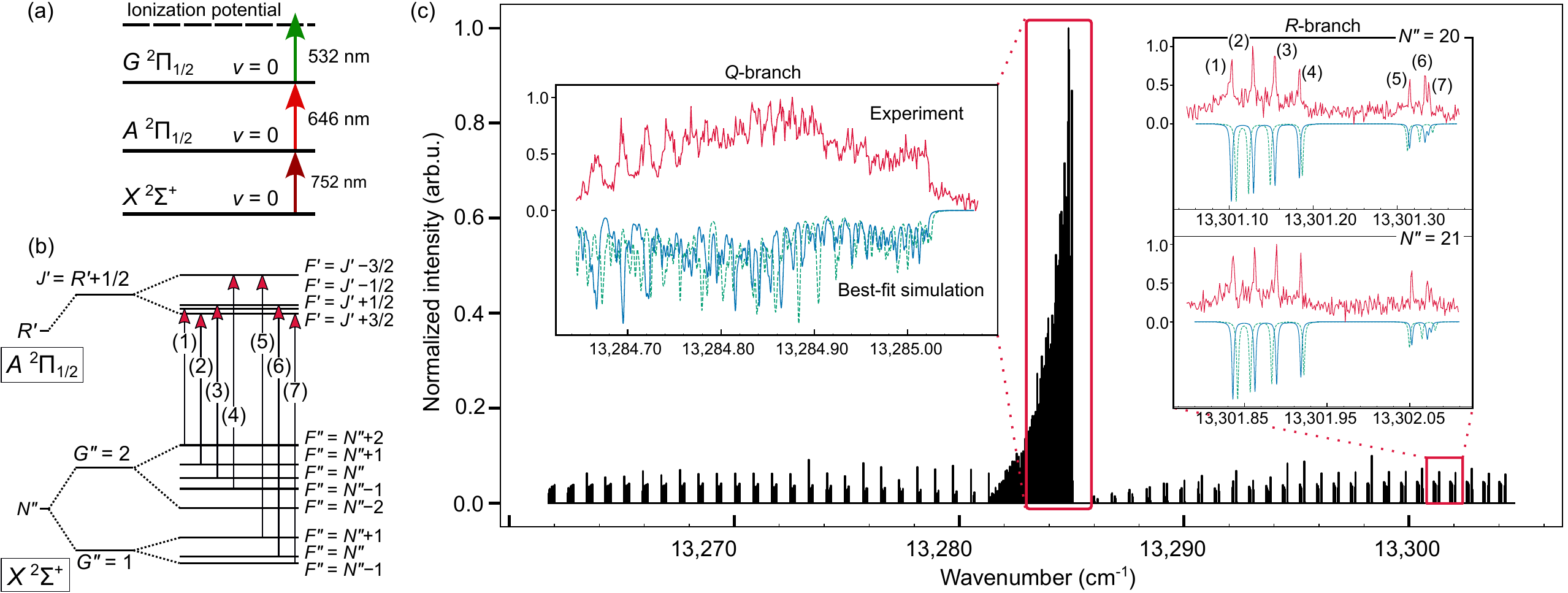}
\caption{(a)~Three-photon resonance ionization scheme used in this work. (b)~Schematic representation of the angular momenta involved in the measured hyperfine peaks for an example transition. (c)~Hyperfine-resolved rotational lines of the $A\,^2\Pi_{1/2}\leftarrow X\,^2\Sigma^+$ transition in $^{223}$Ra$^{19}$F. Inset figures show the measured $Q$-branch and two example $R$-branch rotational lines, where the solid lines represent the best fit and the dashed line is a simulation that neglects the quadrupole hyperfine structure in $X\,^2\Sigma^+$. The $x$-axis shows the Doppler-corrected wavenumber of the 752-nm probe laser, and the $y$-axis shows the count rate on the single-ion detector following background correction.\label{fig:figSpectra}}
\end{figure*}

The fast molecular ion beam was neutralized in flight via collisions with sodium vapor, with residual ions being electrostatically deflected. The neutral $^{223}$Ra$^{19}$F molecules were probed with pulsed lasers in a three-photon laser ionization scheme [Fig.~\ref{fig:figSpectra}(a)]. An injection-seeded titanium-sapphire pulsed laser (1-kHz) at 752\,nm (20-MHz linewidth) excited the molecules from the $X\,^2\Sigma^+$ ground state to the $A\,^2\Pi_{1/2}$ first excited electronic state. A 646-nm pulsed (100-Hz) dye laser (9-GHz linewidth) subsequently excited the molecules to the high-lying $G\,^2\Pi_{1/2}$ state. Finally, a 532-nm Nd:YAG laser (100-Hz) ionized the molecules that had been resonantly excited to $G\,^2\Pi_{1/2}$. The ions were electrostatically separated from the remaining neutral molecules and detected with a single-ion detector.

The hyperfine spectra were collected by tuning the 752-nm laser frequency while monitoring the count rate on the single-ion detector. From the spectroscopic fit, the magnetic dipole and electric quadrupole hyperfine splittings induced by the $^{223}$Ra moments interacting with the molecular fields in the $X\,^2\Sigma^+$ ground state and $A,^2\Pi_{1/2}$ excited state were extracted, along with the molecular rotational constants and their centrifugal distortion constants, the spin-rotation and $\Lambda$-doubling constants, and the term energy of the upper electronic state~\footnote[1]{see Supplemental Material for details}. Similar to Refs.~\cite{udrescu2024precision,conn2025productionspectroscopycoldradioactive}, the predicted dipole hyperfine splitting due to $^{19}$F could not be resolved at the present spectral resolution and was therefore neglected in this analysis.

To extract the nuclear magnetic dipole moment $\mu$ and the spectroscopic electric quadrupole moment $Q_s$ from measured molecular hyperfine splittings, precise calculations of the hyperfine fields within the molecule for each electronic state are required. The precision of the calculated magnetic field has already been demonstrated in Ref.~\cite{wilkins2025observation}, enabling the extraction of the point-like magnetic dipole moment and the magnetization distribution from the $X\,^2\Sigma^+$ state in $^{225}$RaF. Since a single optical transition yields hyperfine constants of both $X\,^2\Sigma^+$ and $A\,^2\Pi_{1/2}$ states, $\mu$ is extracted here from both measurements simultaneously, which yields a value corrected for the Bohr-Weisskopf effect~\footnotemark[1]. Comparing the resulting value with literature provides a test of the employed relativistic coupled-cluster calculations.

For the nuclear electric quadrupole moment, each electronic state has only one dominant contribution to the hyperfine structure, whose form is,
\begin{equation}
    Q_s = \frac{eq_0Q}{V_{zz}},
    \label{eq:Q_eq_0Q}
\end{equation}
where $Q_s$ is the spectroscopic nuclear electric quadrupole moment, $eq_0Q$ is the experimentally measured electric quadrupole molecular hyperfine factor, and $V_{zz}$ is the EFG at the nuclear location in a given electronic state. For $X\,^2\Sigma^+$ in RaF, we use the EFG value $V_{zz}=-1216.6(49)$\,MHz/$e$b calculated in Ref.~\cite{skripnikov2026subpercentaccuracymolecularelectric}. This calculation improves upon the computational treatment presented in Ref.~\cite{wilkins2025observation} by accounting for the effects of finite nuclear structure on the EFG~\cite{koch2010electron} and including the Breit interaction. The obtained value has an overall uncertainty of 0.4\%, achieving a fourfold higher accuracy than the most recent ionic calculations in Ref.~\cite{li2021ab}.

Finally, as shown in Refs.~\cite{athanasakis2023king,Kogel2025BaF}, if the same rovibronic line has been measured in two or more isotopologues of the same molecule, the differences in measured frequency (known as the isotope shift) can be used to extract the difference in mean-squared nuclear charge radius $\delta\langle r^2 \rangle ^{A',A}$ between any two isotopes $A$ and $A'$. 

\paragraph{Results and discussion---}The $Q$-branch band head and six $R$-branch rotational lines of the $A\,^2\Pi_{1/2}$ $\leftarrow$ $X\,^2\Sigma^+$ transition were measured for $^{223}$Ra$^{19}$F. The $R$-branch lines were determined as starting from the rotational quantum numbers $N''=11$--$14$ and $N''=20,21$ in $X\,^2\Sigma^+$ \footnotemark[1].
Each hyperfine peak within each $R$-branch rotational line was fitted with a Voigt lineshape over a polynomial background. The resulting peak centroids and their 1$\sigma$ uncertainties were used to fit the molecular Hamiltonian governing the $X\,^2\Sigma^+$ and $A\,^2\Pi_{1/2}$ states through a bootstrap approach \footnotemark[1]. The resulting hyperfine constants of $^{223}$Ra$^{19}$F are shown in Table~\ref{tab:molecularCts}. The fitted linewidth of the hyperfine peaks ranged from 58(12) to 118(18)\,MHz, marking a nearly three-fold improvement in resolution for the narrowest peak compared to previous spectroscopic measurements using hot, fast beams of RaF at CERN~\cite{udrescu2024precision,wilkins2025observation} while matching the linewidth achieved for RaF and RaOH in a cryogenic buffer gas cell~\cite{conn2025productionspectroscopycoldradioactive}.

\begin{table}[ht]
\caption{\label{tab:molecularCts}Molecular constants (in MHz) of the $X\,^2\Sigma^+$ and $A\,^2\Pi_{1/2}$ electronic states in $^{223}$Ra$^{19}$F. Statistical and systematic uncertainties at the 1$\sigma$ level are given in brackets. The hyperfine constants refer to the $^{223}$Ra nucleus.}
\begin{ruledtabular}
\begin{tabular}{ccc}
Parameter & This work & Theory \\
\hline
\multicolumn{3}{c}{$X^2\Sigma^+$}\\\hline
                 $B''$ & 5,761.914(20)$_{\rm{stat}}$(6)$_{\rm{syst}}$ &  5,753(51) \footnotemark[1]\\
                 $\gamma ''$ & 174.72(20)$_{\rm{stat}}$(4)$_{\rm{syst}}$ & -\\
                 $D''$ & 0.0042(2)$_{\rm{stat}}$(3)$_{\rm{syst}}$\footnotemark[2] & -\\ 
               $b''=A^{\rm{exp}}_{\perp,\Sigma}$ & 2,019.3(15)$_{\rm{stat}}$(4)$_{\rm{syst}}$ & 2,016(50)\footnotemark[3] \\
             $c''$ & $79(4)_{\rm{stat}}$(1)$_{\rm{syst}}$ & 79(72)\footnotemark[3] \\  
             $A^{\rm{exp}}_{\parallel,\Sigma} = b''+c''$ & 2,098.4(35)$_{\rm{stat}}(15)_{\rm{syst}}$ & 2,095(52)\footnotemark[3] \\
             $eq''_0Q$ & $-$1,473(8)$_{\rm{stat}}$(2)$_{\rm{syst}}$ & $-$1,473(82)\footnotemark[3] \\\hline
\multicolumn{3}{c}{$A^2\Pi_{1/2}$}\\\hline
 $T_0$ & 398,265,188(2)$_{\rm{stat}}$(100)$_{\rm{syst}}$ & -\\
                 $B'$ & 5,732.355(20)$_{\rm{stat}}$(6)$_{\rm{syst}}$ & 5,732(51)\footnotemark[1] \\
                 $D'$ & 0.0042(2)$_{\rm{stat}}$(3)$_{\rm{syst}}$\footnotemark[2] & -\\ 
                 $p'$  & $-$12,312.54(11)$_{\rm{stat}}$(3)$_{\rm{syst}}$ & -\\
                 $p_D'$  & 0.0057(6)$_{\rm{stat}}$(20)$_{\rm{syst}}$\footnotemark[2] &  -\\
                 $d'=A^{\rm{exp}}_{\perp,\Pi}$ & 270.1(13)$_{\rm{stat}}$(3)$_{\rm{syst}}$ & 272(7)\footnotemark[3]\\
                 $eq'_0Q$ & $-911(6)_{\rm{stat}}$(2)$_{\rm{syst}}$ & - \\
\end{tabular}
\end{ruledtabular}
\footnotetext[1]{Scaled from Ref.~\cite{athanasakis2025electron}.}
\footnotetext[2]{Fixed to Ref.~\cite{udrescu2024precision}.}
\footnotetext[3]{Taken from Refs.~\cite{skripnikov2020nuclear,petrov2020energy}.}
\end{table}

The hyperfine structure of the $X\,^2\Sigma^+$ state is highly sensitive to the nuclear electric quadrupole moment. Omitting the quadrupole interaction in the simulated spectra shifts the hyperfine components [Fig.~\ref{fig:figSpectra}(b)] of each $R$-branch line by up to 200\,MHz, several times the narrowest observed linewidth [see dashed line in Fig.~\ref{fig:figSpectra}(c)]. As a result, $eq_0''Q=-1473(8)_{\rm stat}(2)_{\rm syst}$\,MHz is determined with a total uncertainty of 0.56\%. 
The systematic uncertainty accounts for drifts and calibration of the ion-beam kinetic energy ($<$2 and $<$0.5\,MHz, respectively) and for fluctuations in the wavemeter readout of the probe laser ($<$1\,MHz), estimated from the continuous measurement of a 780-nm laser locked to a saturation-absorption cell of $^{87}$Rb. An additional absolute systematic uncertainty of 100\,MHz was added to the term energy of the $A\,^2\Pi_{1/2}$ state due to absolute voltage offsets seen in previous CRIS experiments \cite{Lalanne2026}.



\begin{figure}[ht]
    \centering
    \includegraphics[width=\linewidth]{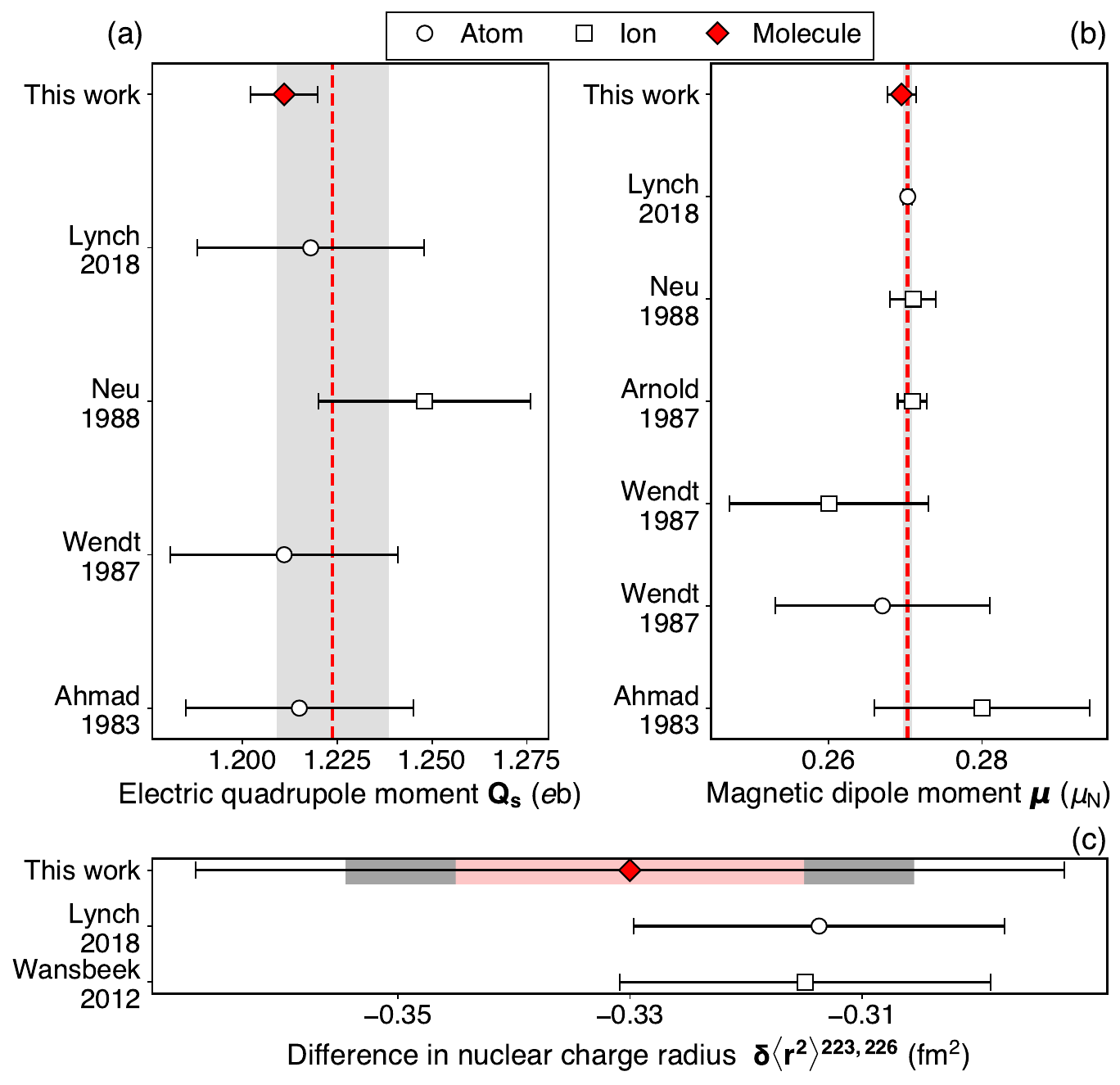}
    \caption{Comparison of the (a) electric quadrupole $Q_s$ and (b) nuclear magnetic dipole $\mu$ moments of $^{223}$Ra, and (c) the difference in mean-square radius $\delta\langle r^2\rangle^{223,226}$, with literature values for each observable~\cite{ahmad1983determination,wendt1987hyperfine,arnold1987direct,neu1988quadrupole,lynch2018laser,wansbeek2012charge}. The red dashed lines in (a) and (b) indicate the error-weighted mean of the literature values for each moment. The shaded areas mark one standard deviation of the weighted mean. In (c), the uncertainty of $\delta \langle r^2 \rangle ^{223,226}$ from this work is dominated by the theory. The outer gray box can be achieved with a threefold reduction in the theoretical error. A further reduction in the systematic error of $T_0$ in $^{226}$Ra$^{19}$F~\cite{udrescu2024precision} down to the level achieved here would result in the uncertainty given by the inner box.} 
    \label{fig:figMoments}
\end{figure}

Similar to other alkaline-earth-like species, the electronic ground states of Ra (total electronic angular momentum $J=0$) and Ra$^+$ ($J=1/2$) are insensitive to $Q_s$, so that quadrupole moments must be extracted from excited states. In molecules, by contrast, the quadrupole interaction is observable in the electronic ground state for any nucleus with $I\geq1$. Moreover, the $X\,^2\Sigma^+$ state of RaF exhibits a large axial EFG at the radium nucleus, resulting in $eq_0''Q= -1473$\,MHz, which is about two times larger than in the most sensitive atomic Ra state measured to date ($7s7p\,^3P_{2}$)~\cite{wendt1987hyperfine}. The quadrupole moment of $^{223}$Ra was thus determined here to be $Q_s = +1.211(7)_{\rm{stat}}(2)_{\rm{syst}}(5)_{\rm{theo}}$\,$e$b, extracted from the measured electric quadrupole hyperfine factor in $X\,^2\Sigma^+$.
Adding the statistical, systematic, and theoretical uncertainties in quadrature gives a total relative uncertainty of 0.73\%.
While past experiments with Ra and Ra$^+$ reported more precise experimental hyperfine factors, the combination of high-accuracy EFG calculations and improved resolution compared to past experiments with short-lived radioactive molecules in this work leads to an approximately three-fold reduction in the uncertainty of $Q_s(^{223}\mathrm{Ra})$ compared to past studies [Fig.~\ref{fig:figMoments}(a)], which are re-evaluated from the hyperfine constants reported in the original works~\cite{ahmad1983determination,wendt1987hyperfine,arnold1987direct,neu1988quadrupole,lynch2018laser} using the latest EFG calculations for the relevant states of Ra and Ra$^+$~\cite{bieron2005degree,li2021ab}. 




The reliability of the underlying electronic-structure calculations used here is supported by the magnetic dipole moment extracted from the same spectra. Using the parallel and perpendicular hyperfine constants of the $X\,^2\Sigma^+$ and $A\,^2\Pi_{1/2}$ states with the electronic factors of Ref.~\cite{skripnikov2020nuclear}, both $\mu$ and the magnetization distribution parameter $B_s$ are determined simultaneously, such that the Bohr-Weisskopf effect is explicitly accounted for in the extracted dipole moment \footnotemark[1]. The resulting $\mu(^{223}\rm{Ra})=+0.2695(17)_{\rm{stat}}(4)_{\rm{syst}}(4)_{\rm{theo}}\,\mu_{\rm N}$ agrees with the atomic and ionic values as shown in Fig.~\ref{fig:figMoments}(b). The most precise magnetic moment value comes from collinear resonance ionization spectroscopy of atomic radium by Lynch \textit{et al.}~\cite{lynch2018laser}, with a precision of 0.2\%. The accuracy of this number depends on the magnitude of the hyperfine anomaly in the studied $^{223}$Ra atomic level. Ahmad \textit{et al.}~\cite{ahmad1983determination} quoted an anomaly of $<0.6\%$ for a similar atomic transition. Despite the atomic uncertainty not including a potential hyperfine anomaly, the magnetic moment from the molecular work shows excellent agreement with that number and all other literature values, giving confidence in the accuracy of the electronic-structure calculations in this work.

\begin{table}[h]
\caption{\label{tab:nuclearPropertiesThisWork} Nuclear magnetic dipole moment $\mu$, magnetization distribution parameter $B_s$, and electric quadrupole moment $Q_s$ of $^{223}$Ra, as well as difference in mean-squared charge radius between $^{223}$Ra and $^{226}$Ra determined from the hyperfine structure of the $X^2\Sigma^+$ and $A^2\Pi_{1/2}$ states in $^{223}$Ra$^{19}$F.
}
\begin{ruledtabular}
\begin{tabular}{lc}
$\mu$ ($\mu_{\rm N}$) & $+0.2695(17)_{\rm{stat}}(4)_{\rm{syst}}(4)_{\rm{theo}}$ \\
$B_s$ ($10^{6}$ MHz/$\mu_{\rm N}$) & $+6.1(10)_{\rm{stat}}(3)_{\rm{syst}}$\footnotemark[1]\\
$Q_s$ ($e$b) &  $+1.211(7)_{\rm{stat}}(2)_{\rm{syst}}(5)_{\rm{theo}}$ \\
$\delta\langle r^2\rangle^{223,226}$ (fm$^2$) &  $-0.33(1)_{\rm{stat}}(2)_{\rm{syst}}(3)_{\rm{theo}}$\\
\end{tabular}
\end{ruledtabular}
\footnotetext[1]{The theoretical error is less than 2\% of the statistical error and has been neglected in this table.}
\end{table}

Lastly, combining the term energy $T_0$ of the $A\,^2\Pi_{1/2}$ state in $^{223}$Ra$^{19}$F and $^{226}$Ra$^{19}$F~\cite{udrescu2024precision} with the calculated field-shift factor~\cite{udrescu2021isotope} yields a $\delta\langle r^2\rangle^{223,226}$ consistent with atomic and ionic results~\cite{wansbeek2012charge} [Fig.~\ref{fig:figMoments}(c)]. Its uncertainty is currently dominated by the field-shift factor and by the systematic uncertainty of $T_0$ in $^{226}$Ra$^{19}$F, both of which can be reduced with improved calculations within the limits of \textit{ab initio} quantum chemistry [outer gray box], and a re-measurement of $^{226}$Ra$^{19}$F at the experimental accuracy achieved here [inner box].



\paragraph{Outlook---}As shown in Table~\ref{tab:nuclearPropertiesThisWork}, molecular spectroscopy can extract nuclear electromagnetic properties of radioactive species with a precision comparable to, or in favorable cases better than, atomic spectroscopy. It can thus be envisioned as an approach to study isotopes that are poorly produced as atomic ions, whose production at RIB facilities is hindered by isobaric contamination, or whose neutral and ionic atomic-level structures are complex or unfavorable.

Light nuclei are a prominent example. The addition or removal of protons and neutrons in these systems gives rise to a wide range of nuclear-structure phenomena, such as halo nuclei~\cite{tanihata1985measurements,jonson1994halo}, nuclear superfluidity and continuum dynamics~\cite{brink2005nuclear,Hagen2012Continuum}, nuclear clustering~\cite{freer2007clustered}, and structures dominated by three-nucleon forces~\cite{Hebeler2015ThreeNucleonForces,Hebeler20213NF}. However, many of these nuclei correspond to reactive elements, such as boron, carbon, nitrogen, oxygen, and fluorine, which are particularly challenging for laser spectroscopy. Molecular spectroscopy instead exploits the natural tendency of these elements to form molecules, which can be delivered as intense, high-purity radioactive ion beams. At ISOLDE, for instance, unstable oxygen isotopes have been extracted with significant yields as $^{12}$C$^{19-22}$O$^+$~\cite{koster2005isol}, a molecule whose hyperfine structure has previously been studied with laser spectroscopy~\cite{de2011uv}, and unstable carbon isotopes as $^{10-15}$C$^{16}$O$^+$~\cite{kronberger2013production,mendonca2014production}. Other highly reactive non-metals, such as phosphorus, sulfur, astatine, and selenium, could be studied in hydride or carbonyl form; radioactive $^{67,68}$Se$^{12}$C$^{16}$O$^+$ beams have already been produced and used for decay spectroscopy of $^{67,68}$Se~\cite{baumann1994beta,koster2003oxide}.

Likewise, laser spectroscopy of unstable halogen isotopes has been severely hindered by their high reactivity and, in the case of iodine, by the lack of optical transitions from the atomic and ionic ground states and of long-lived, high-lying metastable states. These isotopes could instead be studied as aluminum halides, whose production has already been demonstrated~\cite{koster2008progress,hagebo1992new}.

Finally, molecules may provide access to higher-order nuclear moments, such as the magnetic octupole moment, which has so far been measured for only a few nuclei~\cite{fuller1976nuclear,deGroote2021Octupole}. Its hyperfine contribution requires an electronic state with $J>1$, which most atoms and ions with simple level structures lack. In molecules, the rotational angular momentum couples to the electronic angular momentum, so a suitable level can be selected in almost any molecule by choosing a sufficiently high rotational state~\cite{Hammerle1973,Cederberg1999}. Combined with the long interaction times offered by the electronic ground state and the precision of microwave spectroscopy~\cite{BrownCarringtonBook}, this makes molecules a promising platform for such measurements.


\paragraph{Conclusions---}
The spectroscopic electric quadrupole moment of the short-lived nucleus $^{223}$Ra was determined from hyperfine-resolved collinear resonance ionization spectroscopy of $^{223}$Ra$^{19}$F. This constitutes the first molecular determination of the quadrupole moment of a short-lived nucleus produced at an RIB facility, achieving an approximately three-fold reduction in total uncertainty over previous values from atomic and ionic radium. This was enabled by the large sensitivity of the $X\,^2\Sigma^+$ ground state of RaF to the quadrupole hyperfine interaction, combined with a recent high-accuracy relativistic coupled-cluster calculation of the EFG~\cite{skripnikov2026subpercentaccuracymolecularelectric}. The magnetic dipole moment extracted from the same spectra, which explicitly includes the effect of the nuclear magnetization distribution, agrees with atomic measurements, supporting the reliability of the underlying electronic-structure calculations. These results establish molecular spectroscopy as a precision tool for nuclear electromagnetic structure studies on unstable isotopes, complementary to atomic spectroscopy and particularly valuable for elements whose atomic or ionic structure is unfavorable for such measurements.

\paragraph{Acknowledgments---}
We thank the ISOLDE Collaboration and the ISOLDE technical teams for their continued support. This project has received funding from the European Union’s Horizon Europe Research and Innovation programme EUROLABS under Grant Agreement No. 101057511 and Grant Agreement No. 654002 (ENSAR2), the Belgian Excellence of Science (EOS) Project No. 40007501, KU Leuven C1 Project No. C14/22/104, FWO-Vlaanderen Project No. G081422N and 1001323N and International Research Infrastructures (IRI) Project No. I001323N. UKRI STFC Grants No. ST/S002316/1, ST/X00502X/1, ST/P004423/1, ST/V001116/1 and ST/Y000323/1. This project has received funding from the European Union’s Horizon 2020 Research and Innovation Program (grant No. 861198 project ‘LISA’ MSC ITN). K.M. Lynch acknowledges support from the Royal Society Grant No. DHF~\textbackslash R1\textbackslash231007 and UKRI under the Horizon Europe funding Guarantee Grant No. EP/Y036816/1 (ESPEN). K.T. Flanagan acknowledges support from UKRI under the Horizon Europe funding Guarantee Grant No. EP/Z534092/1 (ACTMOL). R.F.G.R. acknowledges support by the Office of Science U.S. Department of Energy, under grants DESC0021176 and DE-SC0026217.

\bibliography{apssamp}

\end{document}


\title{Supplemental material: Precise and accurate nuclear electric quadrupole moment of $^{223}$Ra from molecular spectroscopy}

\author{C.~M.~Fajardo-Zambrano}
\thanks{These authors contributed equally to this work. Contact authors: carlos.fajardo@kuleuven.be, m.athkak@cern.ch}
\affiliation{KU Leuven, Instituut voor Kern- en Stralingsfysica, Leuven, Belgium}

\author{M.~Athanasakis-Kaklamanakis}
\thanks{These authors contributed equally to this work. Contact authors: carlos.fajardo@kuleuven.be, m.athkak@cern.ch}

\affiliation{KU Leuven, Instituut voor Kern- en Stralingsfysica, Leuven, Belgium}
\affiliation{Experimental Physics Department, CERN, Geneva, Switzerland}
\affiliation{JILA and University of Colorado, Boulder, CO 80309, USA}


\author{L.~Lalanne}
\affiliation{Experimental Physics Department, CERN, Geneva, Switzerland}
\affiliation{Université Paris-Saclay, CNRS/IN2P3, IJCLab, Orsay, 91405, France}

\author{J. R. Reilly}
\affiliation{Department of Physics and Astronomy, The University of Manchester, Manchester, UK}

\author{O.~Ahmad}
\affiliation{KU Leuven, Instituut voor Kern- en Stralingsfysica, Leuven, Belgium}

\author{M.~Au}
\affiliation{Systems Department, CERN, Geneva, Switzerland}

\author{S.~W.~Bai}
\affiliation{School of Physics and State Key Laboratory of Nuclear Physics and Technology, Peking University, Beijing, China}

\author{J.~Berbalk}
\affiliation{KU Leuven, Instituut voor Kern- en Stralingsfysica, Leuven, Belgium}
\affiliation{Systems Department, CERN, Geneva, Switzerland}

\author{C.~Bernerd}
\affiliation{Systems Department, CERN, Geneva, Switzerland}

\author{A.~A.~Breier}
\affiliation{Institut f\"ur Physik und Astronomie, Technische Universit\"at Berlin, Berlin, Germany}
\affiliation{Laboratory for Astrophysics, Institute of Physics, University of Kassel, Kassel, Germany}

\author{K.~Chrysalidis}
\affiliation{Systems Department, CERN, Geneva, Switzerland}

\author{T.~E.~Cocolios}
\affiliation{KU Leuven, Instituut voor Kern- en Stralingsfysica, Leuven, Belgium}

\author{R.~P.~de~Groote}
\affiliation{KU Leuven, Instituut voor Kern- en Stralingsfysica, Leuven, Belgium}

\author{K.~T.~Flanagan}
\affiliation{Department of Physics and Astronomy, The University of Manchester, Manchester, UK}
\affiliation{Photon Science Institute, The University of Manchester, Manchester, UK}

\author{S.~Franchoo}
\affiliation{Université Paris-Saclay, CNRS/IN2P3, IJCLab, Orsay, 91405, France}

\author{R.~F.~Garcia~Ruiz}
\affiliation{Department of Physics, Massachusetts Institute of Technology, Cambridge, MA, USA}
\affiliation{Laboratory for Nuclear Science, Massachusetts Institute of Technology and MIT-Harvard Center for Ultracold Atoms, Cambridge, MA, USA}

\author{R.~Heinke}
\affiliation{Systems Department, CERN, Geneva, Switzerland}

\author{P.~Imgram}
\affiliation{KU Leuven, Instituut voor Kern- en Stralingsfysica, Leuven, Belgium}

\author{A.~Koszor\'us}
\affiliation{KU Leuven, Instituut voor Kern- en Stralingsfysica, Leuven, Belgium}

\author{P.~Lass\`egues}
\affiliation{KU Leuven, Instituut voor Kern- en Stralingsfysica, Leuven, Belgium}

\author{J.~Lim}
\affiliation{Centre for Cold Matter, Blackett Laboratory, Imperial College London, London, UK}

\author{Y.~C.~Liu}
\affiliation{School of Physics and State Key Laboratory of Nuclear Physics and Technology, Peking University, Beijing, China}

\author{K.~M.~Lynch}
\affiliation{Department of Physics and Astronomy, The University of Manchester, Manchester, UK}

\author{A.~McGlone}
\affiliation{Department of Physics and Astronomy, The University of Manchester, Manchester, UK}

\author{W.~C.~Mei}
\affiliation{School of Physics and State Key Laboratory of Nuclear Physics and Technology, Peking University, Beijing, China}

\author{G.~Neyens}
 \email{gerda.neyens@kuleuven.be}
\affiliation{KU Leuven, Instituut voor Kern- en Stralingsfysica, Leuven, Belgium}

\author{L.~Nies}
\affiliation{Experimental Physics Department, CERN, Geneva, Switzerland}

\author{A. Raggio}
\affiliation{Department of Physics, University of Jyv\"askyl\"a, Jyväskylä, Finland}

\author{S. Rothe}
\affiliation{Systems Department, CERN, Geneva, Switzerland}

\author{B. van den Borne}
\affiliation{KU Leuven, Instituut voor Kern- en Stralingsfysica, Leuven, Belgium}

\author{J. Warbinek}
\affiliation{GSI Helmholtzzentrum f\"ur Schwerionenforschung GmbH, Darmstadt, Germany}
\affiliation{Department of Chemistry - TRIGA Site, Johannes Gutenberg-Universit\"at Mainz, Mainz, Germany}

\author{J. Wessolek}
\affiliation{Systems Department, CERN, Geneva, Switzerland}
\affiliation{Department of Physics and Astronomy, The University of Manchester, Manchester, UK}

\author{X. F. Yang}
\affiliation{School of Physics and State Key Laboratory of Nuclear Physics and Technology, Peking University, Beijing, China}

\date{\today}

\begin{abstract}
\end{abstract}

\maketitle

\section{Molecular Hamiltonian}
The angular momenta involved in the measured hyperfine transitions are shown schematically in Fig. 1(b) of the main text. Each hyperfine spectrum contains seven hyperfine peaks that arise from transitions between a ground-state rotational level (labeled by its rotational quantum number $N''$) and an excited-state rotational level (labeled by $R'$). These quantum substates in the ground and excited states arise from different angular-momentum couplings. In the $X\,^2\Sigma^+$ state, the intermediate angular momentum $\vec{G}$ with quantum number $G''$ results from the strong coupling of the electron spin, with projection $\Sigma=1/2$, and the nuclear spin of radium, $I=3/2$. $\vec{G}$ then couples with $\vec{N}$ to give the total angular momentum $\vec{F}$ with quantum number $F''$. In the $A\,^2\Pi_{1/2}$ state, $R'$ couples to the projection of the total electronic angular momentum on the internuclear axis, $\Omega=1/2$, yielding $J'$, which then couples to $I$ to give $F'$.

The Hamiltonian fitting was carried out using the command-line version of the publicly available software {\sc pgopher}~\cite{Western2017}. An effective Hamiltonian according to Hund's case (b) was used for the $X\,^2\Sigma^+$ ground state, as,
\begin{equation} \label{eq:XHamiltonian}
    \begin{split}
    H^\Sigma &= H^\Sigma_{\rm{rot}} + H^\Sigma_{\rm{sr}} + H^\Sigma_{\rm{hyperfine}}
    \\
    H^\Sigma_{\rm{rot}} &= B'' \vec{N}^2 - D'' (\vec{N}^2)^2
    \\
    H^\Sigma_{\rm{sr}} &= \gamma'' \vec{N}\cdot\vec{S}
    \\
    H^\Sigma_{\rm{hfs}} &= b'' \vec{I}\cdot\vec{S} + c'' I_zS_z + eq''_0Q\frac{3I_z^2 - \vec{I}^2}{4I(2I-1)} 
    \end{split}
\end{equation}
where $H^\Sigma_{\rm{rot}}$ is the rotational part of the Hamiltonian, $H^\Sigma_{\rm{sr}}$ is the electronic spin-rotation part, and $H^\Sigma_{\rm{hfs}}$ is the hyperfine part. $\vec{N}$ is the total angular momentum excluding spin, $\vec{S}$ is the electronic spin angular momentum, and $\vec{I}$ is the nuclear spin angular momentum. The first term in $H^\Sigma_{\rm{hfs}}$ accounts for the Fermi-contact magnetic dipole interaction, the second term for the magnetic dipole-dipole interaction, and the third term for the electric quadrupole interaction.

For the $A\,^2\Pi_{1/2}$ state, an effective Hamiltonian according to Hund's case (a) was used as,
\begin{equation} \label{eq:AHamiltonian}
    \begin{split}
    H^\Pi &= T_{0} + H^\Pi_{\rm{rot}} + H^\Pi_{\rm{so}} + H^\Pi_{\rm{\Lambda}} + H^\Pi_{\rm{hfs}}
    \\
    H^\Pi_{\rm{rot}} &= B' (\vec{N} - \vec{L})^2 - D' [(\vec{N} - \vec{L})^2]^2
    \\
    H^\Pi_{\rm{so}} &= A' \vec{L}\cdot \vec{S}
    \\
    H^\Pi_{\rm{\Lambda}} &= \frac{1}{2} \left\{ p'+p'_D\vec{J}^2,  ~~e^{2 i \phi} J_{-} S_{-}+e^{-2 i \phi} J_{+} S_{+} \right\}
    \\
    H^\Pi_{\rm{hfs}} &= \frac{1}{2} d' \left(e^{2 i \phi} I_{-} S_{-}+e^{-2 i \phi} I_{+} S_{+}\right) + eq'_0Q\frac{3I_z^2 - \vec{I}^2}{4I(2I-1)} 
    \end{split}
\end{equation}
where $\vec{L}$ is the electronic orbital angular momentum, and $J_{\pm}$, $S_{\pm}$, and $I_{\pm}$ are the raising/lowering operators of the corresponding angular momenta. $T_{0}$ is the excitation energy of $A\,^2\Pi_{1/2}$ (also known as the term energy), $B'$ is the rotational constant and $D'$ the rotational centrifugal distortion constant, $A'$ is the spin-orbit coupling constant, $p'$ is the $\Lambda$-doubling constant, and $p'_D$ its centrifugal distortion constant. In $H^\Pi_{\rm{hfs}}$, $d'$ quantifies the magnetic dipole interaction and $eq'_0Q$ the electric quadrupole interaction.

\section{Laser spectroscopy observables}

\subsection{Magnetic dipole interaction}
In $H^\Sigma_{\rm{hfs}}$ and $H^\Pi_{\rm{hfs}}$, the magnetic dipole interaction is described in terms of the Frosch and Foley parameters ($a,b,c,d$)~\cite{frosch1952magnetic}, which are suited for the interpretation of experimental data using an effective Hamiltonian. In addition to the Frosch and Foley parameters, the coupling strength between the nucleus and the magnetic field produced by the electrons can also be defined as~\cite{haase2020hyperfine,skripnikov2020nuclear,denis2022benchmarking,Chamorro2025BaF},

\begin{equation}\label{eq:abInitio_A}
A_{\nu}=\frac{\mu_n}{I_n S_\nu}\left\langle\sum_i \frac{\left(\vec{r}_{i n} \times \vec{\alpha}_i\right)_\nu}{r_{i n}^3}\right\rangle_{\psi^{(\nu)}}
\end{equation}
where $\mu$ and $I$ are the nuclear magnetic dipole moment and spin of the nucleus. $\nu$ indicates the quantization axis (parallel $\parallel$ or perpendicular $\perp$ to the internuclear axis for diatomic molecules), $S_\nu$ are the projections of the total electronic angular momentum along $\nu$, $\vec{r}_{i}$ is the vector between electron $i$ and the nucleus, $\vec{\alpha}$ are Dirac's matrices, and $\psi$ is the electronic wave function. Because $A_\nu$ can be calculated using \textit{ab initio} quantum chemistry methods, being able to relate $A_{\nu}$ and the Frosch and Foley parameters is important for insightful theory-experiment comparisons. The two sets of parameters are related as~\cite{denis2022benchmarking},

\begin{equation} \label{eq:AnuToFF}
    \begin{split}
    A_{\parallel}^{\Sigma} &= b'' + c''
    \\
    A_{\perp}^{\Sigma} &= b''
    \\
    A_{\perp}^{\Pi} &= d'
    \end{split}
\end{equation}

As shown in \citet{skripnikov2020nuclear}, these parameters can be expressed as,
\begin{equation}\label{eq:magnetic_dipole_expression_1}
    A^{\rm{exp}}_{\parallel/\perp} = A^{(0)}_{\parallel/\perp} + A^{\rm{QED}}_{\parallel/\perp} - A^{\rm{BW}}_{\parallel/\perp}
\end{equation}
where $A^{\rm{exp}}_{\parallel/\perp}$ are the experimentally measured hyperfine factors, $A^{(0)}_{\parallel/\perp}$ are the contributions due to an idealized point-like nuclear magnetic dipole, $A^{\rm{BW}}_{\parallel/\perp}$ correspond to the contributions due to the finite nuclear magnetization (also referred to as the Bohr-Weisskopf effect~\cite{bohr1950influence}), and $A^{\rm{QED}}_{\parallel/\perp}$ account for quantum electrodynamic effects. Separating nuclear and electronic parameters gives,

\begin{equation}\label{eq:magnetic_dipole_expression_3}
\begin{split}
    A^{\rm{exp}}_{\parallel/\perp} &= \frac{\mu}{I} \tilde{A}^{(0)+\rm{QED}}_{\parallel/\perp} - \frac{\mu}{I}B_s\tilde{\tilde{A}}^{\rm{BW}}_{\parallel/\perp}, \\
   \tilde{A}^{(0)+\rm{QED}}_{\parallel/\perp} &= \frac{B^{(0)}_{\parallel/\perp} + B^{\rm{QED}}_{\parallel/\perp}}{\Omega}, \\
   \tilde{\tilde{A}}^{\rm{BW}}_{\parallel} &= \frac{\left( \mathcal{P}_s +\beta \mathcal{P}_p \right)_{\parallel}}{\Omega},\\
   \tilde{\tilde{A}}^{\rm{BW}}_{\perp} &= 2\left( \mathcal{P}_s +\beta \mathcal{P}_p \right)_{\perp}.
\end{split}
\end{equation}

where $\mu$ is the nuclear magnetic dipole moment, $I$ is the nuclear spin, $B_s$ quantifies the radial distribution of nuclear magnetization, $B^{(0)}_{\parallel/\perp} + B^{\rm{QED}}_{\parallel/\perp}$ is the magnetic field due to the electron motion at the location of the nucleus, calculated according to the ideal-dipole picture and including QED corrections, and $\left( \mathcal{P}_s +\beta \mathcal{P}_p \right)_{\parallel/\perp}$ is similarly an electronic factor quantifying the sensitivity of the electronic state to the nuclear magnetization distribution. $\Omega$ is the projection of the total angular momentum along the internuclear axis of the molecule~\cite{skripnikov2020nuclear}.

In Eq.~\eqref{eq:magnetic_dipole_expression_3}, $\tilde{A}^{(0)+\rm{QED}}_{\parallel/\perp} $ and $\dbtilde{A}^{\rm{BW}}_{\parallel/\perp}$ do not depend on the magnetic properties of the nucleus. They are related to the electronic properties of the investigated electronic level in the molecule. High-accuracy values for these electronic factors were calculated in Ref.~\cite{skripnikov2020nuclear} for both the $X\,^2\Sigma^+$ and $A\,^2\Pi_{1/2}$ states in RaF. The point-like and Bohr-Weisskopf electronic parameters for $X\,^2\Sigma^+$ ground state in RaF are,
\begin{equation}\label{eq:magnetic_dipole_theoretical_values_ground}
    \begin{split}
        &\tilde{A}^{(0)+\rm{QED}}_{\parallel} =  12115(121)\,{\rm{MHz}}/\mu_N\\
        &\tilde{A}^{(0)+\rm{QED}}_{\perp} = 11667(117)\,{\rm{MHz}}/\mu_N\\
        &\tilde{\tilde{A}}^{\rm{BW}}_{\parallel} = 7.155(143)\times10^{-5}\,{\rm{MHz}}/\mu_N \\
        &\tilde{\tilde{A}}^{\rm{BW}}_{\perp} = 7.052(141)\times10^{-5}\,{\rm{MHz}}/\mu_N
    \end{split}
\end{equation}

while for the $A\,^2\Pi_{1/2}$ excited state they are,
\begin{equation}\label{eq:magnetic_dipole_theoretical_values_excited}
    \begin{split}
        &\tilde{A}^{(0)+\rm{QED}}_{\parallel} =  1973(20)\,{\rm{MHz}}/\mu_N\\
        &\tilde{A}^{(0)+\rm{QED}}_{\perp} = 1519(15)\,{\rm{MHz}}/\mu_N\\
        &\tilde{\tilde{A}}^{\rm{BW}}_{\parallel} = 4.266(85)\times10^{-6}\,{\rm{MHz}}/\mu_N \\
        &\tilde{\tilde{A}}^{\rm{BW}}_{\perp} = 2.587(52)\times10^{-6}\,{\rm{MHz}}/\mu_N.
    \end{split}
\end{equation}

Taking advantage of the separation between nuclear and electronic parameters in Eq.~\eqref{eq:magnetic_dipole_expression_3}, a system of two equations with two unknowns ($\mu$ and $B_s$) can be constructed using the measured hyperfine factors $A^{\rm{exp}}_{\parallel}$ and $A^{\rm{exp}}_{\perp}$ of the ground and excited state, in our case as,
\begin{equation}\label{eq:2eq2unkown}
\begin{split}
    A^{\rm{exp}}_{\parallel,\Sigma} &= \frac{\mu}{I} \tilde{A}^{(0)+\rm{QED}}_{\parallel,\Sigma} - \frac{\mu}{I}B_s\dbtilde{A}^{\rm{BW}}_{\parallel,\Sigma}, \\
   A^{\rm{exp}}_{\perp,\Pi} &= \frac{\mu}{I} \tilde{A}^{(0)+\rm{QED}}_{\perp,\Pi} - \frac{\mu}{I}B_s\dbtilde{A}^{\rm{BW}}_{\perp,\Pi}.
\end{split}
\end{equation}

In laser spectroscopy, both the parallel and perpendicular hyperfine constants of the upper and lower electronic states in an optical transition are measured simultaneously. Therefore, the set of equations in Eq.~\eqref{eq:2eq2unkown} can be constructed for any isotope by measuring a single optical transition, determining the nuclear observables $\mu$ and $B_s$. This approach can also be applied to atoms and ions~\cite{skripnikov2024reexamination}.

For the extraction of $\mu$ and $B_s$, correlations between $A^{\rm{exp}}_{\parallel}$ and $A^{\rm{exp}}_{\perp}$ are accounted for using a multivariate Gaussian likelihood of the form \cite{sivia2006data},
\begin{equation}
   \begin{split}
        L(\boldsymbol{x})=-\frac{1}{2}\left[(\boldsymbol{d}-\boldsymbol{m(\theta)})^T \boldsymbol{\Sigma}^{-1}(\boldsymbol{d}-\boldsymbol{m(\theta)}) \right. +\\
        \left. \ln (|\boldsymbol{\Sigma}|)+ n\ln (2 \pi) \right]
   \end{split}
\end{equation}

where $\boldsymbol{d} = (A^{\rm{exp}}_{\parallel}, A^{\rm{exp}}_{\perp})$ is the observation vector, $\boldsymbol{m(\theta)}$ is the model prediction, $n$ is the number of observables, and $\boldsymbol{\Sigma}$ is the covariance matrix,
\begin{equation}
    \Sigma=\left(\begin{array}{cc}
\sigma_1^2 & \rho \sigma_1 \sigma_2 \\
\rho \sigma_1 \sigma_2 & \sigma_2^2
\end{array}\right)
\end{equation}

which contains the variances of the individual measurements $(\sigma_1,\sigma_2)$ and their correlation coefficient $\rho$.

\subsection{Electric quadrupole interaction}
For the $X^2\Sigma^+$ ground state,
\begin{equation}
\label{eq:quadrupoleMoment}
        Q_s\,[{\rm{eb}}] =  \frac{eq''_0Q\,[{\rm{MHz}}]}{-1216.6(49)\,[{\rm{MHz/eb}}]}
\end{equation}
where the denominator corresponds to the electric field gradient (EFG) calculated in Ref.~\cite{skripnikov2026subpercentaccuracymolecularelectric}, and $eq_0''Q$ is the measured electric quadrupole hyperfine constant.

Suppl. Table \ref{tab:literatureValue_mu_Q} compares the retrieved nuclear magnetic dipole and electric quadrupole moments of $^{223}$Ra using atomic, ionic, and molecular systems. The reference values are used in Fig.~2 of the main text.

\begin{table}[h]
    \caption{Magnetic dipole $\mu$ and electric quadrupole $Q_s$ moments of $^{223}$Ra as well as difference in mean-squared charge radii $\delta\langle r^2\rangle^{223,226}$ between $^{223}$Ra and $^{226}$Ra in literature.}
\begin{ruledtabular}
    \begin{tabular}{c|cc}
        System &  $\mu$ ($\mu_N$) & Reference \\\hline
        RaF & $+0.2695(17)_{\rm{stat}}(4)_{\rm{syst}}(4)_{\rm{theo}}$ &  This work \\
        Ra & $+0.2703(6)$ &  \cite{lynch2018laser} \\
        Ra & $+0.267(14)$ & \cite{wendt1987hyperfine} \\
        Ra$^+$ & $+0.271(3)$ & \cite{neu1988quadrupole} \\
        Ra$^+$ & $+0.2705(19)$ & \cite{arnold1987direct} \\
        Ra$^+$ & $+0.260(13)$ & \cite{wendt1987hyperfine} \\
        Ra$^+$ & $+0.280(14)$ & \cite{ahmad1983determination} \\\hline
         &  $Q$ ($e$b) &  \\\hline
        RaF & $+1.211(7)_{\rm{stat}}(2)_{\rm{syst}}(5)_{\rm{theo}}$ &  This work \\
        Ra & $+1.218(30)$ & \cite{lynch2018laser,bieron2005degree} \\
        Ra & $+1.211(30)$ & \cite{wendt1987hyperfine,bieron2005degree} \\
        Ra & $+1.215(30)$ & \cite{ahmad1983determination,bieron2005degree}\\
        Ra$^+$ & $+1.248(28)$ & \cite{neu1988quadrupole,li2021ab} \\\hline
         &  $\delta\langle r^2\rangle^{223,226}$ (fm$^2$) &  \\\hline
        RaF & $-0.33(1)_{\rm{stat}}(2)_{\rm{syst}}(3)_{\rm{theo}}$ &  This work \\
        Ra & $-0.31(2)$\footnotemark[1] &  \cite{lynch2018laser} \\
        Ra$^+$ & $-0.31(2)$\footnotemark[2] &  \cite{wansbeek2012charge} \\
    \end{tabular}
    \end{ruledtabular}
    \footnotetext[1]{Re-calculated with $^{226}$Ra as the reference isotope using $F = -28.75(144)$\,GHz and $K= -305(177)$\,GHz\,amu.}
    \footnotetext[2]{Re-calculated with $^{226}$Ra as the reference isotope using $F = -39.81(199)$\,GHz and $K= -990(351)$\,GHz\,amu.}

    \label{tab:literatureValue_mu_Q}
\end{table}

\subsection{Charge radii}
The governing expression, taken from atomic spectroscopy~\cite{king2013isotope}, is 
\begin{equation}
    \label{eq:kingPlot}
    \delta \nu^{A,A'} = F \delta\left\langle r^2 \right\rangle^{A, A'} + K \frac{m_{A'}-m_{A}}{m_{A'} m_{A}},
\end{equation}
where $\delta \nu^{A,A'}$ is the isotope shift of the electronic transition between isotopes $A$ and $A'$, $\delta\left\langle r^2 \right\rangle^{A, A'}$ is the difference in their mean-squared charge radii, $m_{A}$ and $m_{A'}$ are the (atomic or ionic) masses of the two isotopes, and $F$ and $K$ are the electronic field-shift and mass-shift factors, respectively, which are specific to the electronic transition measured but independent of the specific isotopes. A similar expression can be derived for molecules \cite{athanasakis2023king}.

The extracted $\delta \langle r^2 \rangle ^{223,226}$ is taken from the isotope shift $\delta \nu^{223,226}_{T_0}$ in $T_0$ between $^{223,226}$Ra$^{19}$F \cite{udrescu2024precision}. Due to the large mass of the Ra isotopes, the mass shift ($< 30$ MHz) is smaller than the systematic uncertainty on the absolute frequency measurements $T_0$ between experiments, and has therefore been neglected. Thus, the difference in charge radii was extracted with the $F= -24732(2473)$\,MHz factor taken from Ref.~\cite{udrescu2021isotope} as,
\begin{equation}
    \delta \langle r^2 \rangle ^{223,226} = \frac{\delta \nu^{223,226}_{T_0}}{F}.
    \label{eq:chargeRadii}
\end{equation}

\begin{figure*}[ht]
\includegraphics[width=\textwidth]{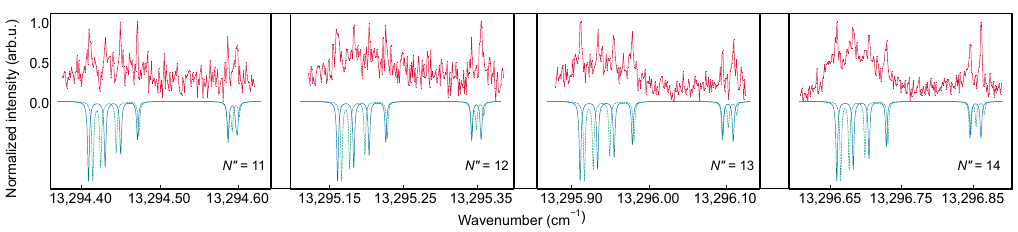}
\caption{Hyperfine-resolved $R$-branch rotational lines of the $A\,^2\Pi_{1/2}\leftarrow X\,^2\Sigma^+$ transition in $^{223}$Ra$^{19}$F starting from $N''=11$--$14$, complementing the two lines shown in Fig.~1(c) of the main text. Solid lines show the spectral simulation based on the best-fit Hamiltonian, and dashed lines show a simulation neglecting the quadrupole hyperfine structure of $X\,^2\Sigma^+$. The $x$-axis shows the Doppler-corrected wavenumber of the 752-nm probe laser, and the $y$-axis shows the count rate on the single-ion detector following background correction. \label{fig:figSpectra}}
\end{figure*}

\section{Extended methods}
The data analysis was performed in 4 steps. In step 1, the well-resolved hyperfine spectra from the R-branch (see Fig.~1 in the main text and Suppl. Fig.~\ref{fig:figSpectra}) were fitted to determine the centroids and full width at half-maximum (FWHM) of the observed peaks. These centroids were then used in step 2, where first (approximate) values for the molecular and hyperfine parameters of $^{223}$Ra$^{19}$F were deduced by fitting the R-branch, ignoring the Q-branch, using the appropriate molecular Hamiltonian for each electronic state in {\sc pgopher}. The resulting parameters were then used in step 3 to determine which of the many peaks in the Q-branch are resolved and belong to a specific hyperfine spectrum. Nineteen such lines could be identified. In the 4th and final step, all data were fit together using a bootstrap approach to determine the final molecular constants and hyperfine parameters.

\subsection{Line-shape analysis}
\begin{figure}[ht]
    \centering
    \includegraphics[width=0.9\linewidth]{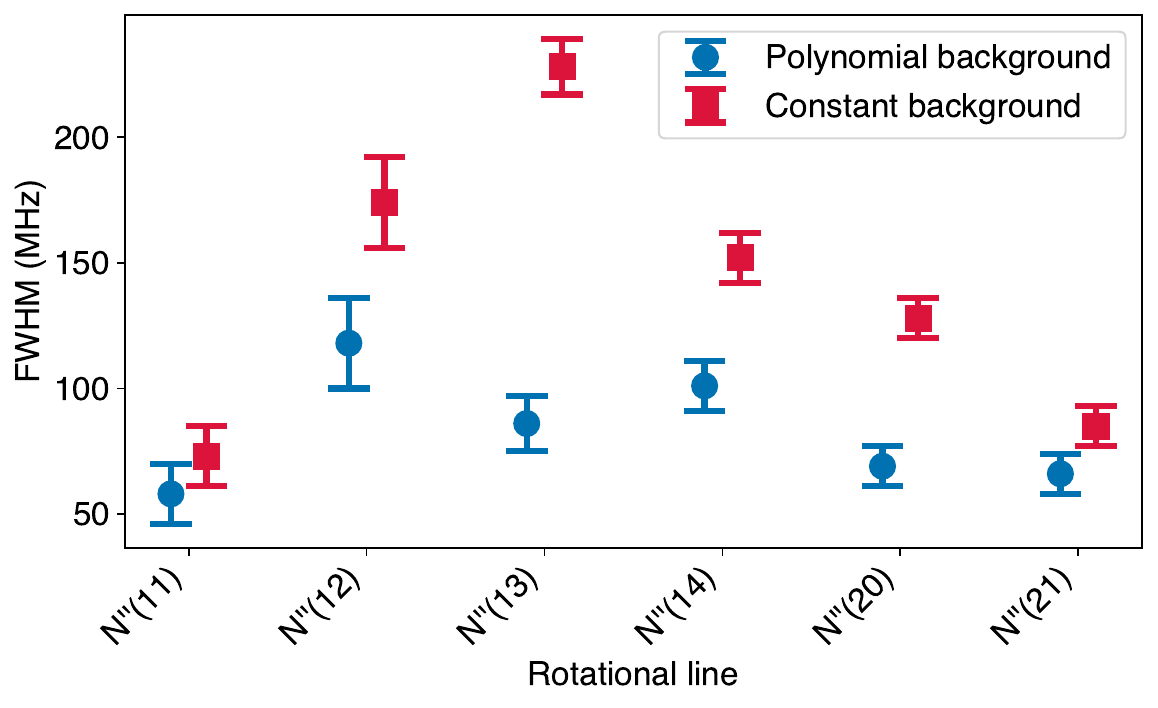}
    \caption{Comparison of the full width at half-maximum (FWHM) of each rotational line using a polynomial and constant background.}
    \label{fig:FWHM}
\end{figure}

A total of six well-resolved hyperfine spectra were measured in this experiment, characterized by the ground-state rotational quantum number $N''$. Each hyperfine spectrum was initially fitted with seven independent Voigt profiles over a second-order polynomial background using the SATLAS package \cite{GINS2024109053}. The centroid of each peak in each hyperfine spectrum has been extracted, with the FWHM of all peaks within a hyperfine multiplet constrained to be equal. This polynomial background was chosen after visual inspection of the observed line shapes in the $R$-branch, shown in the main Fig.~1(c) and Suppl. Fig.~\ref{fig:figSpectra}. To cross-check this choice, the line shapes were also obtained using a constant background. Suppl. Table~\ref{tab:FWHM} shows the fitted FWHM for each rotational line with constant and polynomial background models. Under the constant background model, the FWHM varies across the different rotational lines, despite measurements being performed under similar conditions. In contrast, the polynomial background leads to a more consistent FWHM (see Suppl. Fig.~\ref{fig:FWHM}). Therefore, the polynomial background was used to fit the hyperfine structures, and the resulting centroids were used in further analysis. 

\begin{table}[ht]
\caption{\label{tab:FWHM} Full width half maximum of each measured rotational line using a Voigt distribution over a constant and second-order polynomial background.}
\begin{ruledtabular}
\begin{tabular}{ccc}
 Rotational line & FWHM$_{\rm{cte}}$ (MHz) & FWHM$_{\rm{poly}}$ (MHz) \\\hline
$N''(11)$ & 73(11) & 58(12) \\
$N''(12)$ & 174(22) &  118(18) \\
$N''(13)$ & 228(20) & 86(11) \\
$N''(14)$ & 152(13) &  101(10) \\
$N''(20)$ & 128(11) &  69(8) \\
$N''(21)$ & 85(8) & 66(8)\\
\end{tabular}
\end{ruledtabular}
\end{table}

\vspace{-0.5cm}

\begin{figure*}[ht]
    \centering
    \includegraphics[width=0.75\linewidth]{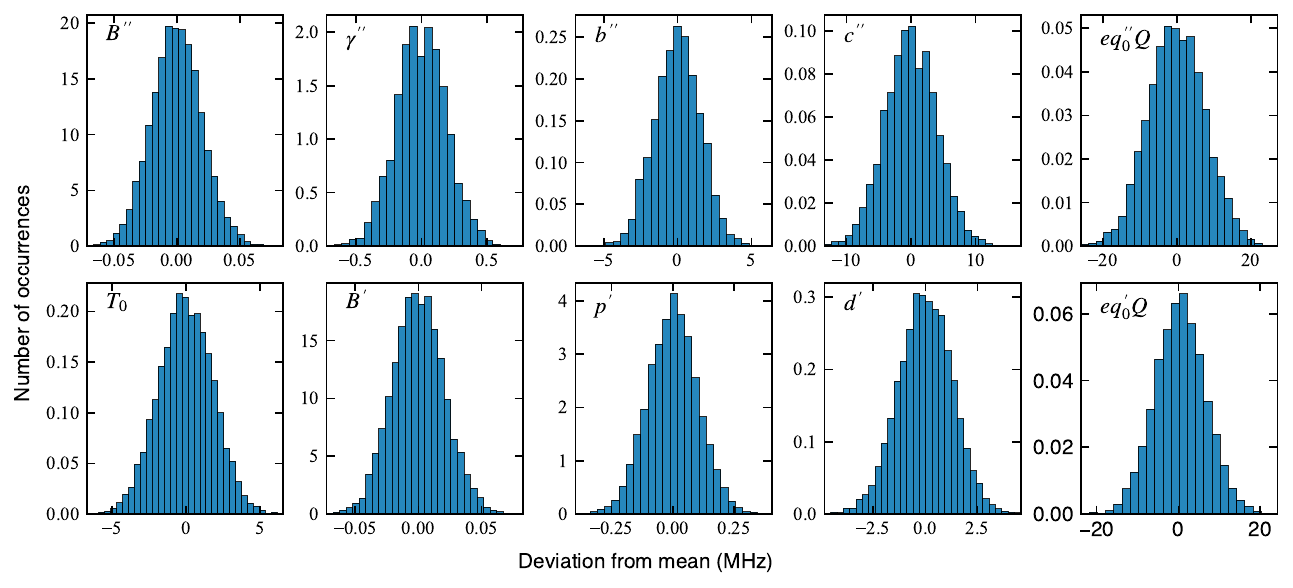}
    \caption{Distributions of fitted molecular constants after 10,000 bootstrap iterations. The mean of each distribution is subtracted for visualization purposes.}
    \label{fig:histogramMolCts}
\end{figure*}

\subsection{Data analysis}
The $Q$-branch is highly congested with rotational lines, each of which is further split by hyperfine interactions with the $^{223}$Ra nucleus. This produces a dense spectrum in which each peak in Fig.~1(c) of the main text is composed of multiple overlapping hyperfine peaks, hindering its assignment. Thus, the initial fitting of the molecular spectra included only the $R$-branch lines, neglecting the $Q$-branch. For this, the initial guess of the rotational and centrifugal distortion constants $B'',B',D''$, and $D'$ of the $X\,^{2}\Sigma^+$ and $A\,^{2}\Pi_{1/2}$ states were scaled from the $^{226}$Ra$^{19}$F values~\cite{udrescu2024precision} using the square root of the ratio of the reduced masses of the two isotopologues~\cite{gerhard1950molecular}. Similarly, the initial guesses for the magnetic hyperfine factors $b'',c'',d'$ in $^{223}$Ra$^{19}$F were scaled from the values measured for $^{225}$Ra$^{19}$F~\cite{skripnikov2020nuclear,wilkins2025observation}, based on literature values for the nuclear magnetic dipole moments of $^{223}$Ra and $^{225}$Ra ~\cite{lynch2018laser}.

The molecular constants obtained by fitting the six R-branch hyperfine spectra with {\sc pgopher} were used to identify nineteen distinct (resolved) hyperfine peaks in the Q-branch. The assignment of their quantum numbers used the transition-strength data, which were also provided by {\sc pgopher}. The centroids of these nineteen peaks and their assigned quantum numbers were then used in a combined fit, together with the six hyperfine spectra from the R-branch, to produce the final rotational and hyperfine constants, as given in the main Table 1. A bootstrap approach was used to determine these values and their errors, as outlined in the next sections.



\subsection{Bootstrap estimates}
\begin{figure*}[ht]
    \centering
    \includegraphics[width=\linewidth]{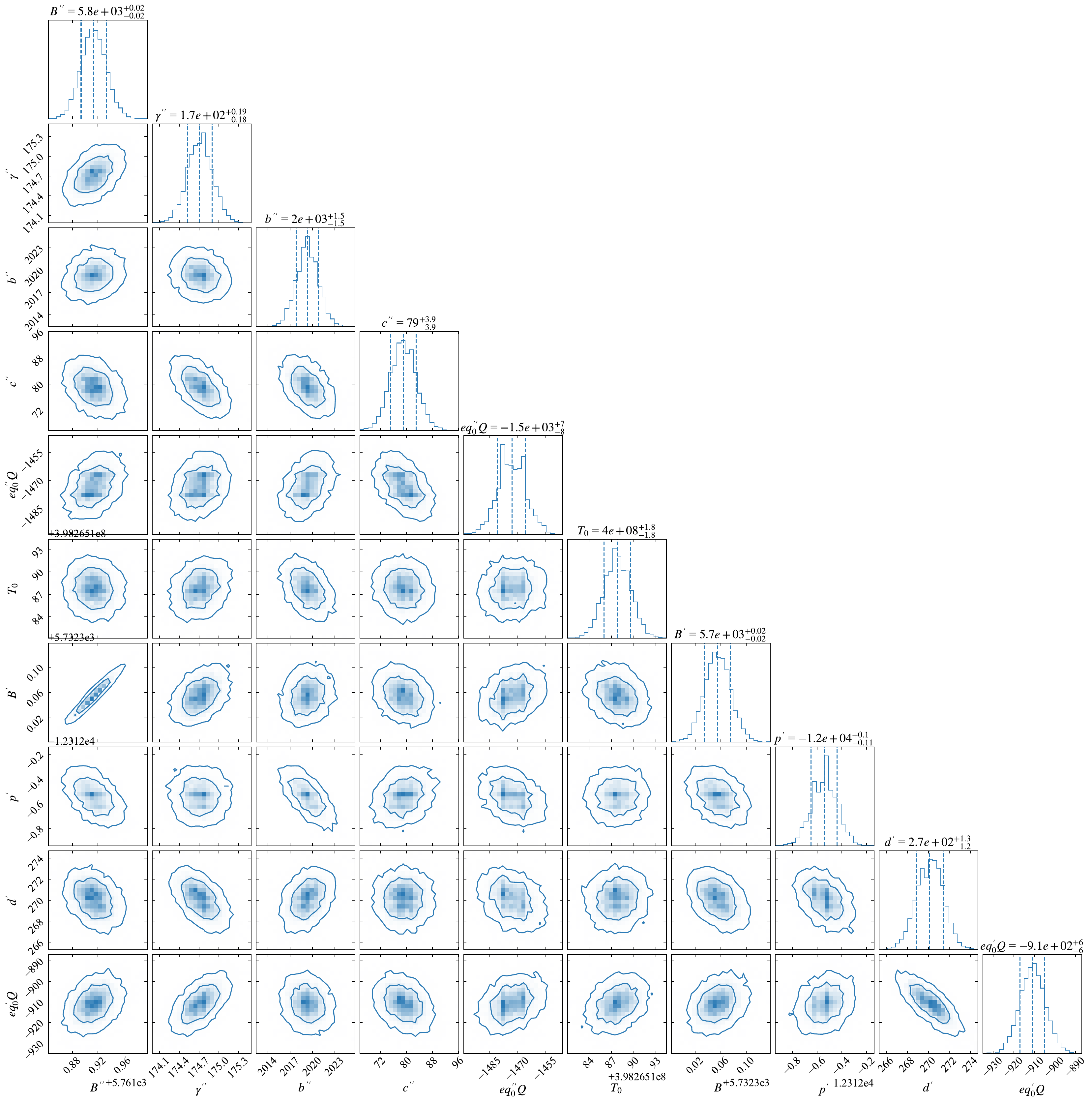}
    \caption{Ten-dimensional marginal distributions for each molecular constant included in the fit of $^{223}$RaF.}
    \label{fig:CornerMolCts}
\end{figure*}

The nominal values and uncertainties of the molecular constants were obtained with a bootstrap resampling procedure. Each hyperfine peak within the $R$-branch lines and the nineteen identified $Q$-branch peaks was represented by a Gaussian distribution, with the peak's best-fit centroid and 1$\sigma$ uncertainty from the Voigt-lineshape analysis as the mean and standard deviation. In each bootstrap iteration, the molecular Hamiltonian was fitted to a set of peak centroids randomly sampled from these distributions, and the resulting molecular constants were stored. After 10,000 iterations, the mean of each constant's distribution was taken as its nominal value, and the standard deviation as its 1$\sigma$ statistical uncertainty (Suppl. Fig.~\ref{fig:histogramMolCts}). To obtain the systematic uncertainties, the procedure was repeated with the width of each peak's Gaussian distribution replaced by the linear sum of all known systematic error sources. The marginal distributions are Gaussian and unimodal, supporting the reported uncertainties, and they show no evidence of multimodality or significant nonlinear correlations. Most parameters are only weakly correlated, with the notable exception of the rotational constants (Suppl. Fig.~\ref{fig:CornerMolCts}); these correlations are accounted for in the extraction of the magnetic dipole moment.

\subsection{$Q$-branch convergence} 
\begin{figure*}[ht]
    \centering
    \includegraphics[width=\linewidth]{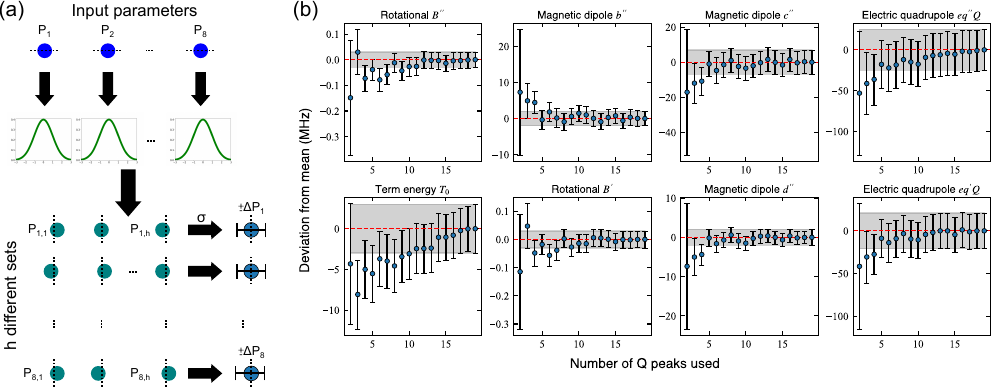}
    \caption{(a) Bootstrap scheme showing the procedure used to extract the nominal values and uncertainties of the hyperfine molecular constants. (b) Difference between the final values for the hyperfine molecular constants and the best-fit results obtained for different numbers of $Q$-branch transitions included in the fit.}
    \label{fig:figBootstrap}
\end{figure*}

As aforementioned, the assignment of the Q-branch was performed based on an initial fit of the molecular constants, allowing the identification of nineteen distinct peaks that primarily consist of a single hyperfine transition in the $Q$-branch of $^{223}$Ra$^{19}$F. Although this assignment was guided by the transition-strength data provided by {\sc pgopher}, a Q-branch convergence test was implemented to ensure the data were unbiased. Accordingly, the hyperfine structure was fitted repeatedly using random combinations of between two and eighteen of these assigned $Q$-branch transitions to avoid biasing the final fit. This procedure was repeated thirty times, and the average results, along with the weighted mean and standard deviation $\sigma$, are shown in Suppl. Fig. \ref{fig:figBootstrap}(b). All extracted hyperfine coupling constants agreed with the final results as long as at least five $Q$-branch transitions were included in the fit, regardless of which transitions were chosen.

\subsection{$R$-branch convergence}
\begin{figure*}[ht]
    \centering
    \includegraphics[width=0.7\linewidth]{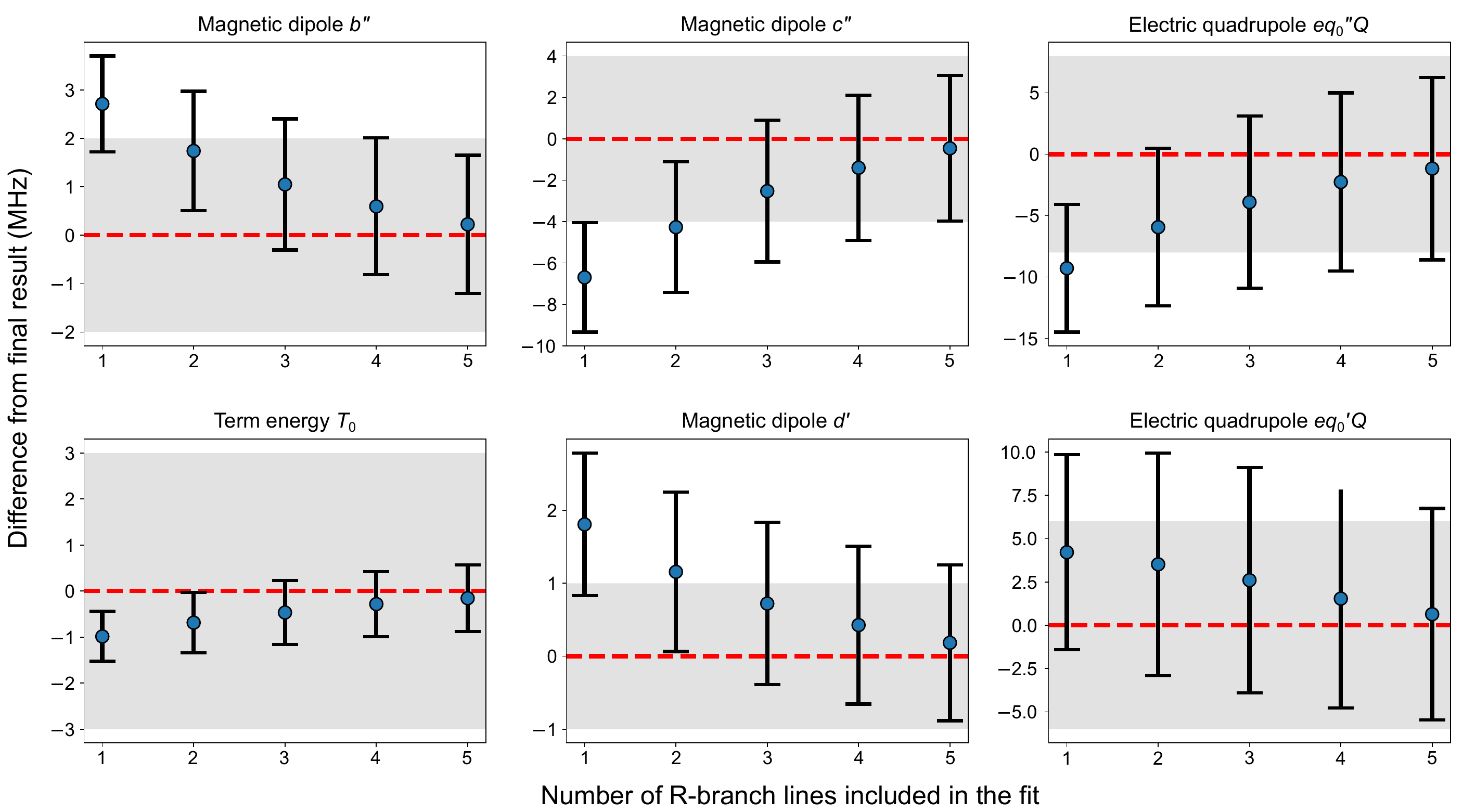}
    \caption{Change in the fitted hyperfine molecular constants based on the number of $R$-branch rotational lines included in the fit. Upper row: Constants for the $X\,^2\Sigma^+$ state. Lower row: Constants for the $A\,^2\Pi_{1/2}$ state. All $Q$-branch peaks included in the final fit were also included in all fitting iterations shown in this plot.}
    \label{fig:R-convergance}
\end{figure*}

As previously mentioned, six rotational lines of the $R$-branch were measured in this experiment. The $R$-branch lines measured were $^{s}R_{1G}(11)$, $^{s}R_{1G}(12)$, $^{s}R_{1G}(13)$, $^{s}R_{1G}(14)$, $^{s}R_{1G}(20)$ and $^{s}R_{1G}(21)$, where $G=1,2$ for each line, following the spectroscopic notation of Ref.~\cite{le2009hyperfine}. 
As for the $Q$-branch, it is important to determine how many $R$-branch lines are required to obtain reliable molecular constants. To this end, the spectrum was refitted using subsets of two to five of the six measured $R$-branch lines, covering all possible combinations of lines. For example, two lines can be selected in fifteen different ways from the six, including both adjacent and widely separated pairs. The nineteen assigned $Q$-branch peaks were included in every one of these fits, since the fit did not converge on $R$-branch lines alone. The molecular constants retrieved from each subset were then compared with the final result obtained using all six $R$-branch lines and the $Q$-branch. Suppl. Fig.~\ref{fig:R-convergance} shows how the retrieved hyperfine constants vary as the number of $R$-branch lines included in the fit increases.

Evidently, all molecular constants were within one standard deviation of the final result (obtained using all six $R$-branch rotational lines) if at least three $R$-branch rotational lines were included. In addition, regardless of the number of rotational lines included, the retrieved coupling constants were closer to the final result in fittings where the rotational lines were most energetically separated than in sets that used successive rotational lines. 

\clearpage
\bibliography{apssamp}